\documentclass[reprint,amssymb,amsmath,aip,cha,frontmatterverbose]{revtex4-2}
\usepackage{graphicx}
\usepackage{dcolumn}
\usepackage{bm}
\usepackage{subfigure}

\begin{document}

\title{Patched patterns and emergence of chaotic interfaces in arrays of nonlocally coupled excitable systems}%

\author{Igor Franovi\'c}
\email{franovic@ipb.ac.rs}
\affiliation{Scientific Computing Laboratory, Center for the Study of Complex Systems,
Institute of Physics Belgrade, University of Belgrade, Pregrevica 118, 11080 Belgrade, Serbia}
\author{Sebastian Eydam}
\email{richard.eydam@riken.jp}
\affiliation{Neural Circuits and Computations Unit, RIKEN Center for Brain Science, 2-1 Hirosawa, 351-0198 Wako, Japan}

\date{\today}

\begin{abstract}
We disclose a new class of patterns, called patched patterns, in arrays of non-locally coupled excitable units with attractive and repulsive interactions. Self-organization process involves formation of two types of patches, majority and minority ones, characterized by uniform average spiking frequencies. Patched patterns may be temporally periodic, quasiperiodic or chaotic, whereby chaotic patterns may further develop interfaces comprised of units with average frequencies in between those of majority and minority patches. Using chaos and bifurcation theory, we demonstrate that chaos typically emerges via a torus breakup and identify the secondary bifurcation that gives rise to chaotic interfaces. It is shown that the maximal Lyapunov exponent of chaotic patched patterns does not decay, but rather converges to a finite value with system size. Patched patterns with a smaller wavenumber may exhibit diffusive motion of chaotic interfaces, similar to
that of the incoherent part of chimeras.
\end{abstract}

\maketitle

\begin{quotation}
While coherence-incoherence patterns have been exhaustively explored both for spatially discrete systems of coupled oscillators and in the continuum limit, much less is known about the generic mechanisms of onset and the finite-size effects associated with such patterns in coupled excitable systems. Recently discovered supercritical scenario for the onset of bumps in coupled excitable active rotators with nonlocal attraction and global repulsion, as well as the two types of solitary states unveiled in arrays of excitable FitzHugh-Nagumo units with nonlocal attractive and repulsive interactions, already suggest that the coherence-incoherence patterns in coupled excitable system may defy the common intuition based on coupled oscillators. Here we introduce a new class of patterns, called patched patterns, in non-locally coupled arrays of excitable units with attractive and repulsive interactions. These patterns involve splitting of an array into spatially continuous domains, called patches, comprised of units locked by their average spiking frequencies.
Patched patterns may be temporally periodic, quasiperiodic or chaotic, and depending on the prevalence of attraction vs repulsion, chaotic patterns can develop interfaces with frequencies intermediate between the majority and minority patches. Distinct from chimeras, chaos in patched patterns is not spatially localized, but is of different character for the units in the patches and at the interfaces: the latter show more variability and engage in chaotic itinerancy. We demonstrate the typical bifurcation scenario giving rise to chaos for smaller and intermediate coupling ranges. We also show that adjusting the coupling range to reduce the pattern wavenumber may result in a transition to chaos accompanied by a diffusive motion of interfaces.
\end{quotation}

\section{Introduction} \label{intro}

Combining different approaches and methods from pattern formation, finite dimensional chaos and bifurcation theory, as well as statistical physics, has within recent years allowed for some deep insights into the coherence-incoherence patterns in systems of coupled oscillators. The two most important aspects concern understanding their mechanisms of onset and the dependence of their behavior on system size. For example, it has become clear that chimeras \cite{KB02,AS04, AMSW08,Z20,PJAS21,PA15,OK19,H21} constitute inhomogeneous equilibria of certain macroscopic averaged quantities in the continuum limit \cite{O18,L11,O13}, while in spatially discrete systems, they are characterized by a self-localized, spatially extensive weak hyperchaos where the positive part of the Lyapunov spectrum decays to zero with system size \cite{WOYM11}. The interplay of local nonlinearity and interactions, at least for the two classical scenarios admitting chimeras \cite{KB02,AMSW08}, results in nontrivial finite-size effects, such as the pattern collapse to a uniform coherent state \cite{WO11} and the Brownian-like diffusion of the incoherent domain \cite{OWM10}.

Nevertheless, an intriguing question is what happens to coherence-incoherence patterns if a system is not comprised of oscillators, but rather of excitable units \cite{LGNS04,I07,FYEBW20,FTPVB15,FTVB12}. When isolated, excitable systems remain in a stable stationary state, but may be triggered to oscillate by a sufficiently strong perturbation via interactions and/or noise. Excitability is one of the corner stones for the physics of life, underpinning the local dynamics of neuronal, cardiac and endocrine systems \cite{I07,AB13,S21,KS09}, and is also important for understanding of many other natural and synthetic systems, from chemical reactions \cite{MS06} and climate dynamics \cite{ABCR21} to lasers \cite{T21} and machine learning \cite{CAL19}. Self-organization in coupled excitable systems cannot be described in terms of a simple mutual adjustment of local oscillations, and even the very onset of a collective mode requires the presence of inhibitory/repulsive couplings \cite{RZ21,RZ21a}, defying the common intuition developed for coupled oscillators. With the full analogy to coupled oscillators missing, the basic questions on coherence-incoherence patterns in coupled excitable systems, such as the potential working definition, classification, generic mechanisms of onset and the contribution from finite-size effects still remain open.

Currently, it seems likely that the extension of the concept of weak chimeras \cite{AB15}, classically associated with small systems of coupled oscillators, provides an effective framework for characterization of coherence-incoherence patterns in finite systems of coupled excitable units. By this concept, coherence-incoherence patterns can be described in terms of frequency locking/unlocking, typically involving a majority of units that are coherent, i.e. frequency locked, and a minority of units unlocked from the bulk (and possibly mutually unlocked). In these terms, bump states \cite{OLC07,L11,L16,LO20}, a common type of pattern associated in neuroscience with spatial working memory \cite{B12} and the head direction system \cite{Z96}, can be described as a set of units with an elevated firing rate self-localized on a continuous spatial domain and appearing on top of an inactive background. Using the model of excitable active rotators with a short-range attraction and long-range repulsion, it has recently been shown that the bumps may emerge from Turing patterns
via a supercritical scenario that involves unlocking of a single unit from the bulk and a subsequent sequence of bifurcations to a fully developed extensive chaos \cite{FOW21}. Such an onset mechanism turned out to be completely independent on system size. In contrast to classical chimeras, no pattern collapse was observed in small systems, and while typically being static, bumps could also exhibit a lateral diffusive motion depending on the parameters. Also quite recently, applying the model of an array of FitzHugh-Nagumo units with nonlocal attractive and repulsive interactions, it has been shown that coupled excitable systems may display two types of solitary states \cite{FESZ22} with a different finite-size behavior, namely either size-independent periodic solutions closely associated with unbalanced cluster states in globally coupled networks, or weakly chaotic solutions where a few isolated oscillators split off from the background alternating (modulated travelling) wave. Finally, for the same model, it has been shown that the noise may play a facilitatory role allowing for the onset of the so-called coherence-resonance chimeras \cite{SZAS16, ZSAS17}, whereby instead of the diffusion drift of classical chimeras, the interplay of local noise and the intrinsic noise due to finite size gives rise to switching of positions between the coherent and the incoherent domain.

In the present paper, we introduce a new class of patterns in non-locally coupled excitable systems, called \emph{patched patterns}. The basic pattern structure is such that the units self-organize into spatially continuous domains, called patches, comprised of units that are mutually frequency locked. Our model is the same as in \cite{OOHS13,FESZ22} and comprises an array of $N$ non-locally coupled identical FitzHugh-Nagumo units described by
\begin{align}
\varepsilon \dot{u}_k&=u_k-\frac{u_k^3}{3}-v_k +\frac{\kappa}{2R}\sum\limits_{l=k-R}^{k+R}
[g_{uu}(u_l-u_k)\nonumber \\
&+g_{uv}(v_l-v_k)]  \nonumber \\
\dot{v_k}&= u_k+ a +\frac{\kappa}{2R}\sum\limits_{l=k-R}^{k+R}
[g_{vu}(u_l-u_k)+g_{vv}(v_l-v_k)] \nonumber \\
\end{align}
The local slow-fast dynamics is paradigmatic of type II excitability \cite{I07}, and involves activator variables $u_k$ and recovery variables $v_k$ with a timescale separation $\varepsilon=0.05$. For an isolated unit, variation of the bifurcation parameter $a>0$ gives rise to a singular Hopf bifurcation at $a=1$, mediating between excitable ($a>1$) and oscillatory regimes ($a<1$). Above the canard transition at $a\approx 1-\varepsilon/8$ \cite{BE86}, harmonic subthreshold (low-amplitude) oscillations give way to large-amplitude relaxation oscillations. Here we fix $a=1.01$, such that the isolated units are in the excitable regime. Each unit is coupled to $R$ nearest neighbors to its left and to its right, with all the indices being periodic modulo $N$. Coupling strength $\kappa$ is assumed to be homogeneous, and is fixed to $\kappa=0.4$. Interactions between units involve direct terms including only activator or only recovery variables, as well as the mixing terms, which is compactly described by the rotational coupling matrix \cite{OOHS13} $G = \left(\begin{matrix}g_{uu} & g_{uv}\\ g_{vu} & g_{vv}\end{matrix}\right)=\left(\begin{matrix}\cos \varphi & \sin \varphi\\ -\sin \varphi & \cos \varphi\end{matrix}\right)$. Note that the parameter $\varphi$ impacts the prevalence of attractive vs repulsive interactions \cite{FESZ22}.

The paper is organized as follows. In Sec. \ref{patt} we first make a basic description of patched patterns, and then focus on static chaotic patched patterns with interfaces to characterize the local switching dynamics of interface units, showing that it consists of laminar and turbulent epochs consistent with chaotic itinerancy \cite{KT03,T09}. In Sec. \ref{chaos} we use chaos and bifurcation theory to demonstrate the typical scenario for the onset of chaos with increasing of the coupling parameter $\varphi$, where the torus bifurcation mediates the transition from periodic to quasiperiodic patterns, and chaos emerges via torus breakup. It is also shown that the observed chaos is extensive and that the maximal Lyapunov exponent converges to a finite value rather than decaying with the system size. In Sec. \ref{diff} we demonstrate how varying the coupling range to reduce the pattern wavenumber may give rise to the diffusion of interfaces.

\section{Patched patterns} \label{patt}

\begin{figure}[ht]
\centering
\includegraphics[scale=0.175]{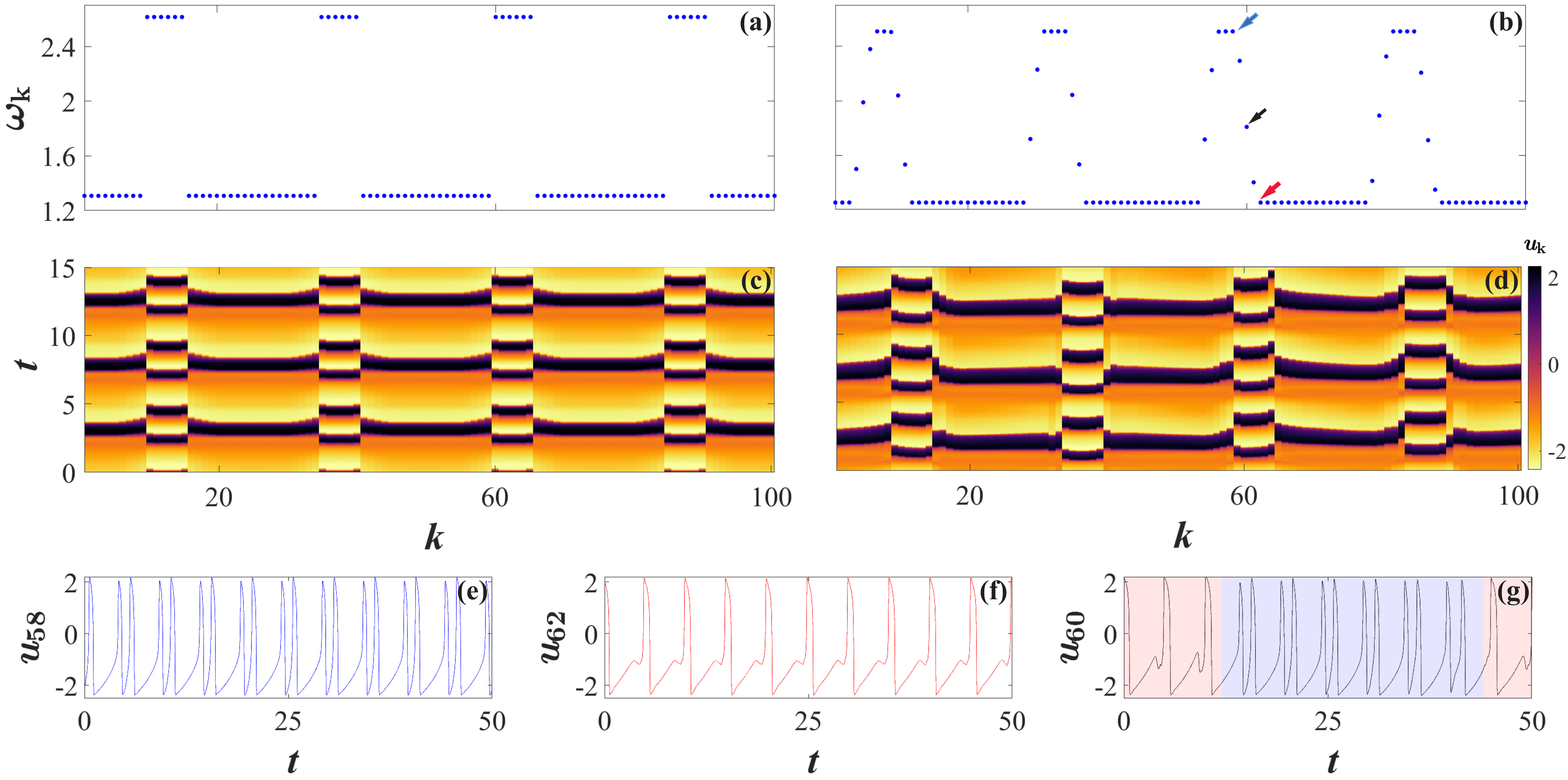}
\caption{Patched patterns without and with chaotic interfaces. (a) and (c): spatial profile of average spiking frequencies and spatiotemporal evolution of fast variables $u_k(t)$ for a periodic patched pattern at $\varphi=2$; (b) and (d) show the same, but for a chaotic patched pattern with interfaces ($\varphi=2.2$). (e), (f) and (g): typical time traces $u_k(t)$ of units from a majority patch ($k=62$, red arrow in (b)), minority patch ($k=58$, blue arrow), and from the interface ($k=60$, black arrow). Red and blue shading in (g) indicate transient episodes where the interface unit attaches to one of the patches. Remaining parameters: $N=100$, $a=1.01$, $\varepsilon=0.05$, $\kappa=0.4$, $R=20$. The time horizon used to obtain average spiking frequencies is $T=10^{6}$ t.u.}
\label{fig1}
\end{figure}

As already announced, patched patterns involve formation of coherent spatial domains of frequency locked units. One may distinguish between two types of domains, here called majority and minority patches. The majority patches maintain a 1:2 resonant frequency locking to the minority patches. Patched patterns can be temporally periodic, quasi-periodic or chaotic. The basic spatial profile of average spiking frequencies $\omega_k=2\pi M_k/T$, where $M_k$ is the spike count within a macroscopic time interval $T$, is piecewise constant, as in Fig.~\ref{fig1}(a), which illustrates a periodic solution with an additional reflection symmetry. In terms of local dynamics, the patches are heterogeneous, such that the units closer to the patch center show a more similar dynamics than those at the patches’ boundaries. In contrast to periodic solutions, the chaotic solutions may further develop \emph{interfaces} comprised of incoherent units with \emph{switching} dynamics, whose frequencies are intermediate between majority and minority patches, see Fig.~\ref{fig1}(b). Depending on the system parameters controlling the pattern wavenumber, i.e. the number of minority patches, these interfaces may be static, as in Fig.~\ref{fig1}(b), or may display Brownian-like diffusive motion we analyze later on in the paper.

\subsection{Switching dynamics at the interfaces} \label{switch}

Let us analyze in more detail the self-organization of local dynamics for an example of a static chaotic pattern with interfaces, whose spatial profile of average spiking frequencies is illustrated in Fig.~\ref{fig1}(b). Such wavenumber-4 pattern emerges from the corresponding periodic solution with a piecewise constant profile of average frequencies, illustrated in Figs. \ref{fig1}(a) and (c), via a sequence of bifurcations described in Sec. \ref{chaos}. The typical time series of a fast variable of a majority unit $k=62$ (see the red arrow in Fig.~\ref{fig1}(b)) indicates mixed mode oscillations with each pair of successive spikes separated by a subthreshold oscillation, whereas the time trace of a typical minority unit $k=58$, denoted by the blue arrow in Fig.~\ref{fig1}(b), primarily shows successive spiking, cf. Fig.~\ref{fig1}(e) and (f). At the other hand, a short time trace of an interface unit $k=60$ in Fig.~\ref{fig1}(g) indicates mixed-mode oscillations with a switching dynamics between the episodes where it approaches either the neighboring majority or the minority patch.

\begin{figure}[ht]
\centering
\includegraphics[scale=0.175]{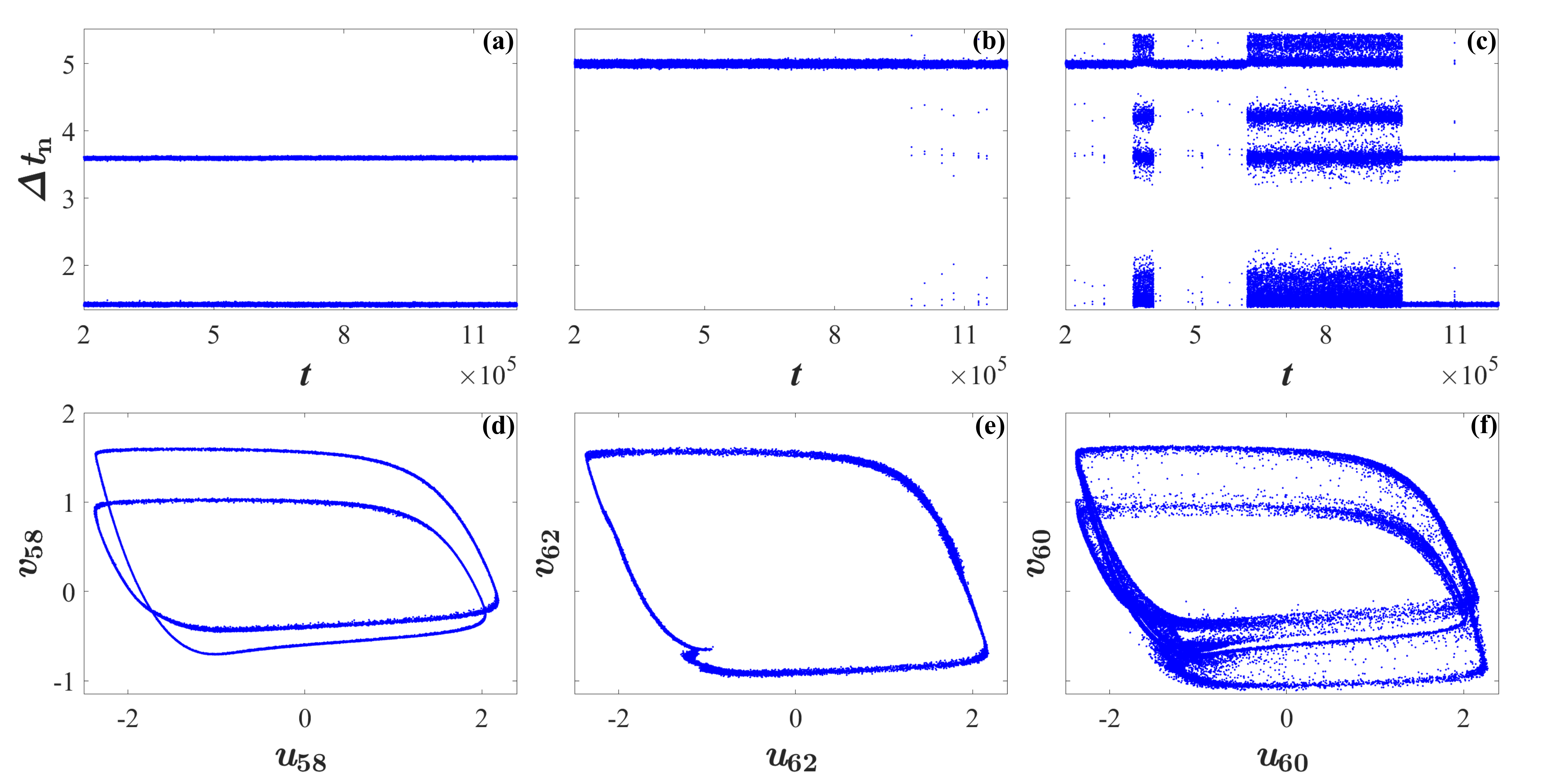}
\caption{Local fluctuations within patches and at interfaces. (a)-(c) Temporal evolution of the first return times to the Poincare cross-section $u_k(t)=1.5,\dot{u}_k(t)>0$ for the representative minority, majority and interface units from Fig.~\ref{fig1}(b), respectively. (d)-(f) Corresponding phase portraits in the $u_k-v_k$ plane. System parameters are the same as in Fig.~\ref{fig1}(b).}
\label{fig2}
\end{figure}

To further elucidate the switching dynamics at the interfaces, we construct the diagrams comparing the time evolution of the first return times $\Delta t_n(t)$ to the Poincare cross-section $u_k(t)=1.5, \dot{u}_k(t)>0$ for the representative majority, minority and interface units from Fig.~\ref{fig1}, see Figs. \ref{fig2}(a)-(c). For a minority unit, one typically observes only small variations around two basic levels, which are just induced by fluctuations of the local mean-field, also see the phase portrait in Fig.~\ref{fig2}(d). The similar holds for the representative majority unit, cf. Figs. \ref{fig2}(b) and (e), though here one also finds larger fluctuations in the first return times derived from rare subthreshold excitations. The most peculiar behavior is manifested by the representative interface unit in Figs. \ref{fig2}(c) and (f), where the dependence $\Delta t_n(t)$ involves a slow alternation between two types of epochs: the laminar ones, when the unit is approximately frequency locked to the closest majority or minority patch, and turbulent ones, when the unit displays a high variability due to fast fluctuations between the orbits resembling those of units in majority and minority patches. Slow alternating dynamics is a hallmark of chaotic itinerancy, ubiquitous in high-dimensional state spaces. In particular, the interface unit visits ruins of localized chaotic attractors resembling the local dynamics within the patches, interspersed with a higher-dimensional irregular motion providing for the switching paths between the ruins. 

\begin{figure}[ht]
\centering
\includegraphics[scale=0.205]{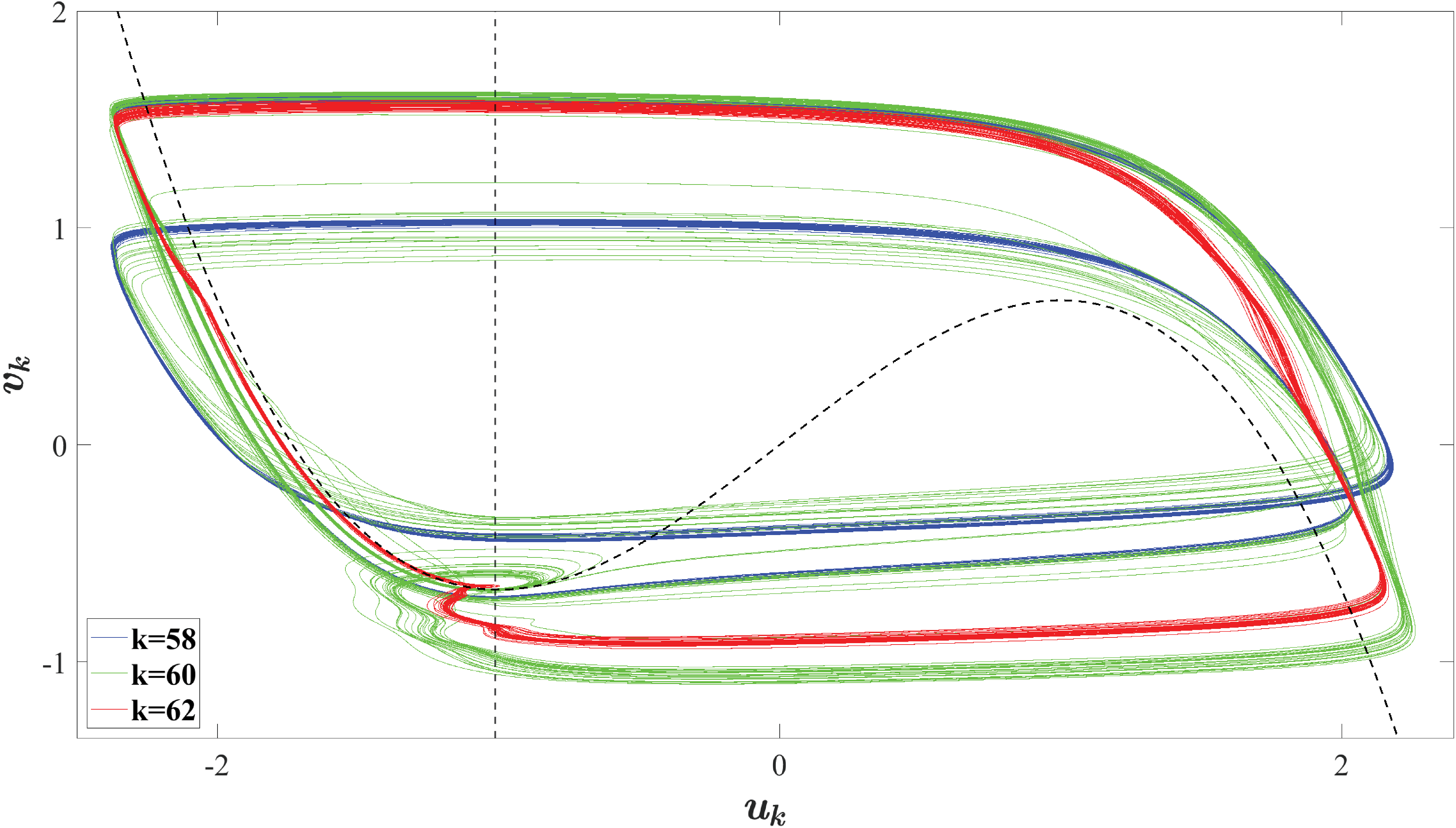}
\caption{Superimposed orbits of a representative minority unit ($k=58$, blue), majority unit ($k=62$, red)   and an interface unit ($k=60$, green) for the pattern in Fig.~\ref{fig1}(b). Black dashed lines: isolated unit’s fast and slow nullclines, whose intersection determines the position of the corresponding stable equilibrium. Fast switching between episodes within the turbulent epoch of an interface unit is due to subthreshold oscillations around the ghost of the isolated unit’s equilibrium.}
\label{fig3}
\end{figure}

The mechanism giving rise to switching between the epochs, as well as the fast fluctuations between the episodes within turbulent epochs, appears to be qualitatively the same. It is associated with the interface unit performing small-amplitude oscillations around the ghost of an unstable fixed point derived from the stable equilibrium of an isolated unit, as illustrated in Fig.~\ref{fig3} for the fast fluctuations within a turbulent epoch. Successive subthreshold oscillations are also the reason of why the turbulent epochs contain $\Delta t_n$ levels absent in case of units within the patches, see Fig.~\ref{fig2}(c) . Relaxation oscillations both within laminar and turbulent epochs are susceptible to perturbations in a way similar to the phenomenon of phase-sensitive excitability of a limit cycle \cite{FOW18, EFW19}, in the sense that a strong enough perturbation due to fluctuations in the local mean-field may induce a large deviation from the relaxation oscillation orbit, giving rise to one or more successive subthreshold oscillations.

\begin{figure}[ht]
\centering
\includegraphics[scale=0.35]{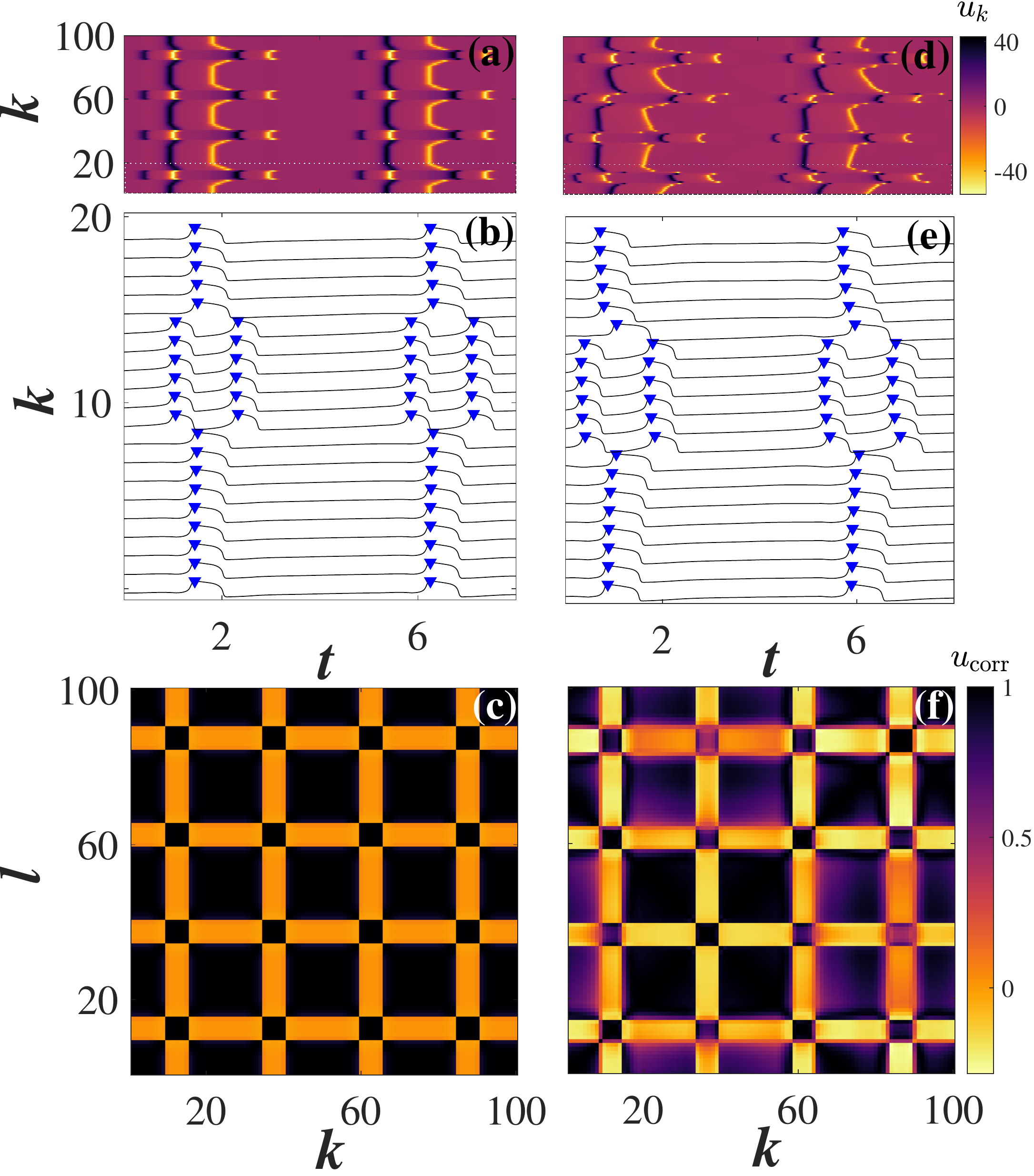}
\caption{Intrinsic structure of typical periodic (left column) and chaotic pattern with interfaces (right column). (a)
Spatiotemporal evolution $\dot{u}_k(t)$; white
dotted rectangle: segment of an array whose dynamics is extracted in (b); (b) bottom-up: time traces $u_k(t),k=1,2,...,20$ (black lines) shown shifted by a constant increment; blue triangles: spike times of units; (c) cross-correlation matrix $C_{kl}$; (d)-(f) Same as (a)-(c) but for chaotic pattern with interfaces. Respective parameters are identical
to those in Figs. \ref{fig1}(a) and \ref{fig1}(b).}
\label{fig4}
\end{figure}

Pattern formation is based on two self-organization mechanisms classically observed in coupled excitable systems, namely \emph{self-localized excitations}\cite{WOS15} and \emph{propagation of excitation}\cite{SZAS16,NG96}. The activity within an array consists of sequentially repeating excitation episodes, where the majority (minority) patches fire once (twice). Within the patches, spiking is typically organized in such a way that the units closer to the center fire before those at the patches’ boundaries. The excitation episodes are initiated at the minority patches, see Fig.~\ref{fig4}(a) and Fig.~\ref{fig4}(d) that show the space-time evolution of $\dot{u}_k,k\in[1,N]$ for the periodic and the chaotic patched pattern with interfaces from Figs. \ref{fig1}(a) and \ref{fig1}(b), respectively. For the periodic pattern, the excitation of the minority patches, see e.g. black regions for $t\approx 1$ occurs simultaneously as the solution carries a reflection space-time symmetry. Contrasting that, the reflection symmetry is broken for the chaotic pattern, cf. Fig.~\ref{fig4}(d). The localized excitation elicited within a minority patch becomes a source of two counterpropagating excitation waves emanating to its left and right. Each majority patch is embedded between two minority patches, and hence receives from them two counterpropagating excitation waves that collide and annihilate. In their wake, another excitation is born, and induces by the described paradigm a second spike of units within the minority patches. The latter cannot induce further excitation at the majority patches because the units there feature longer spikes and subsequently also have longer refractory periods, see Figs. \ref{fig4}(b) and \ref{fig4}(e).

The temporal organization of activity within and between the patches may further be examined by constructing the corresponding cross-correlation matrix $C_{kl}=\frac{\langle \hat{u}_k(t)\hat{u}_l(t) \rangle_T}{\sqrt{\langle \hat{u}_k(t)^2 \rangle_T\langle \hat{u}_l(t)^2 \rangle_T}}$, where $\langle \cdot \rangle_T$ denotes the time averaging, while $\hat{u}_k(t)=u_k(t)-\langle u_k(t)\rangle_T$ are the deviations of $u_k(t)$ from their means, see Fig.~\ref{fig4}(c) and \ref{fig4}(f). One immediately realizes that $C_{kl}$ for the periodic pattern has a clear-cut structure with a strong correlation within and between a given type of patches (majority or minority), while the correlation of activities between patches of different type is rather weak. The intrinsic structure of $C_{kl}$ for the chaotic pattern is more smeared, reflecting the existence of interface units, and in contrast to the periodic solution, there are also pairs of units with an anti-correlated behavior.

\section{Emergence of chaos} \label{chaos}

Having explained the structure of local dynamics underpinning chaotic patched patterns with interfaces, we investigate the bifurcation scenario that gives rise to chaos as the coupling parameter $\varphi$ is increased. Note that the features of the transition to chaos with $\varphi$ depend on the wavenumber of the primary periodic solution, which is ultimately controlled by the coupling range $R$. We first elaborate on a generic scenario where periodic patterns follow the route to chaos via quasiperiodicity, focusing on the example of a wavenumber-4 pattern. For this generic scenario, which holds for smaller and intermediate coupling ranges $R$, the onset of chaos \emph{per se} is not immediately associated with the formation of turbulent interfaces, and the latter emerge separately via a secondary bifurcation on a chaotic attractor.

\begin{figure}[ht]
\centering
\includegraphics[scale=0.17]{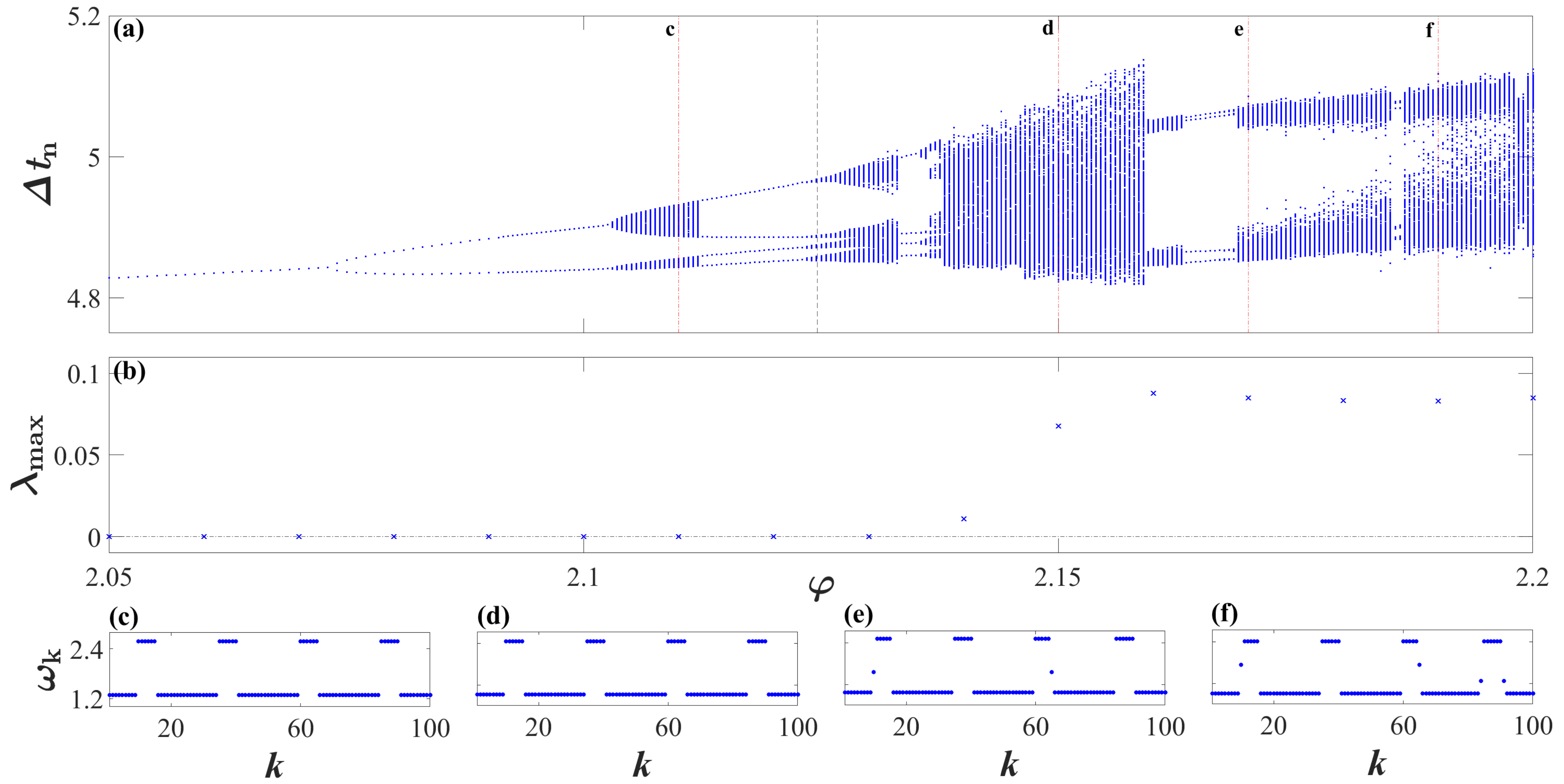}
\caption{Emergence of chaos and formation of turbulent interfaces. (a) Bifurcation diagram of local dynamics: first return times $\Delta t_n$ to the cross-section $u_k(t)=1.5,\dot{u}_k(t)>0$ in dependence of $\varphi$ for the unit $k=45$; black dash-dotted line: $\varphi$ value where the solution loses reflection symmetry. (b) Variation of the maximal Lyapunov exponent $\lambda_{max}$ with $\varphi$. (c)-(f) Spatial profiles of average local spiking frequencies $\omega_k$ for the set of $\varphi$ values indicated by the red dashed lines in (a). Remaining parameters are the same as in Fig.~\ref{fig1}.}
\label{fig5}
\end{figure}

The bifurcation diagram in Fig.~\ref{fig5}(a) is constructed considering an array of $N=100$ units, performing a forward sweep in $\varphi$ to collect the first return times $\Delta t_{n}$ of local dynamics to the Poincare cross-section $u_k(t)=1.5, \dot{u}_k(t)>0$. In the provided example, the selected unit remains within one of the majority frequency patches over the whole considered $\varphi$ interval. The red dashed lines indicate the $\varphi$ values associated with the spatial profiles of average spiking frequencies $\omega_k$ in Figs. \ref{fig5}(c)-(f). The latter are calculated by averaging over an interval $T=10^6$ t.u. having discarded a transient of additional $5\times 10^5$ t.u. The bifurcation diagram is supplemented by the dependence of the maximal Lyapunov exponent $\lambda_{max}(\varphi)$ \cite{PP16,BGS76}, see Fig.~\ref{fig5}(b), sampled for the solutions observed at a fixed increment $\Delta \varphi=0.01$.

\begin{figure}[ht]
\centering
\includegraphics[scale=0.3]{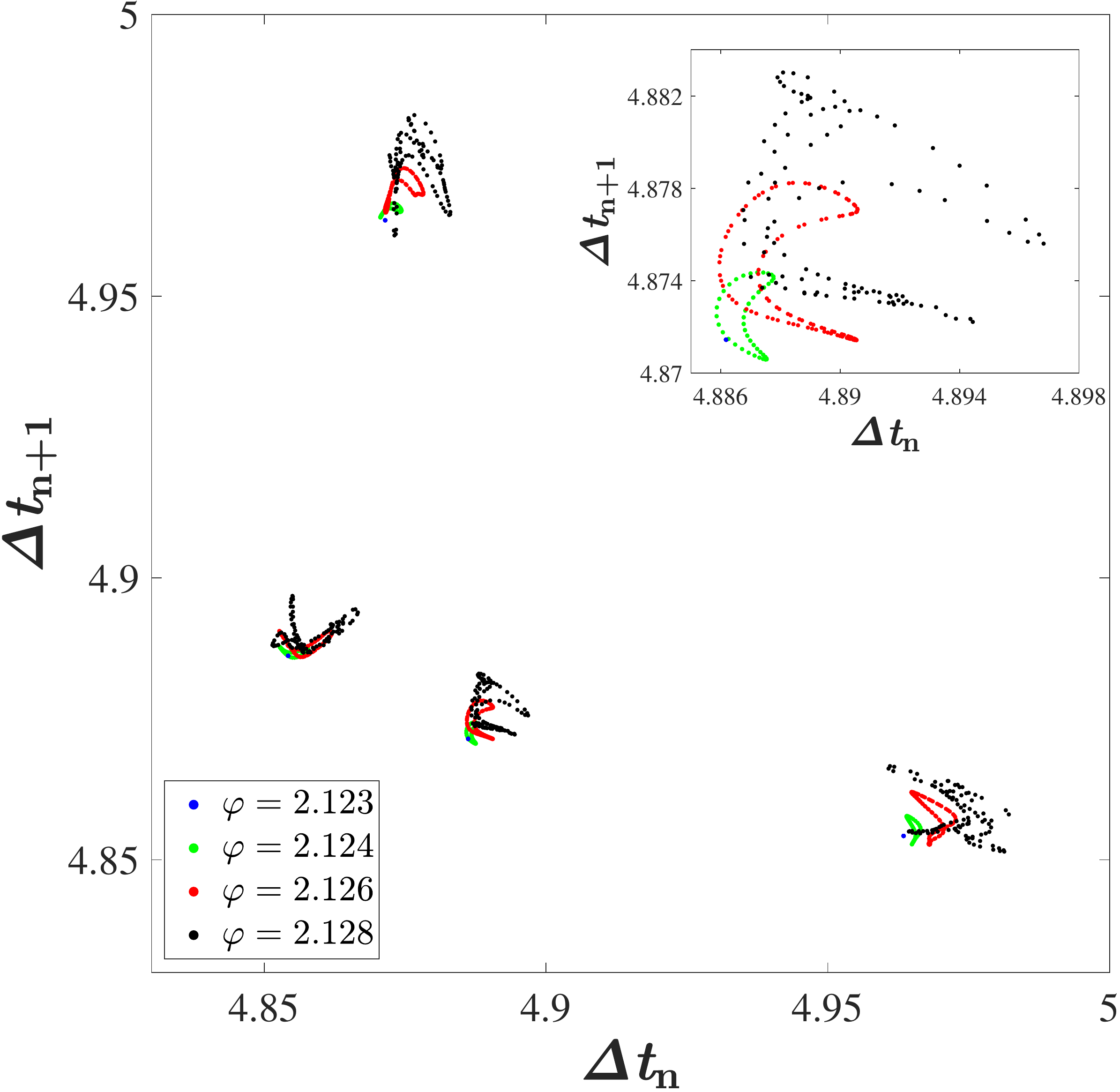}
\caption{Focus on the breakup of torus bifurcation: first return maps $\Delta t_{n+1}(\Delta t_n)$ indicate the disappearance of an invariant curve with increasing $\varphi$. Inset: enlarged view of one of the segments of the relative period-four orbits in the Poincare section. Remaining  parameters are the same as in Fig.~\ref{fig1}.}
\label{fig6}
\end{figure}

The initial state at $\varphi=2.0$ is the periodic patched pattern with a reflection symmetry, already illustrated in Fig.~\ref{fig1}(a) and (c). Following a period doubling at $\varphi\approx 2.073$, the period two pattern is transformed into a stable quasiperiodic solution via a torus bifurcation at $\varphi\approx 2.103$. Further increasing $\varphi$, there is a locking on the torus at $\varphi\approx 2.112$, which is followed by a subsequent transition to chaos via a \emph{torus breakup} around $\varphi\approx 2.128$. The primary pattern, corresponding to a relative periodic orbit with the period four in the Poincare section, as well as the incipient chaotic pattern, are illustrated by the respective first return maps $\Delta t_{n+1}(\Delta t_n)$ in Fig.~\ref{fig6}. For $\varphi\approx 2.125$ just below the transition, cf. the black dash-dotted line in Fig.~\ref{fig5}(a), the solution loses the reflection symmetry, which we have verified by calculating the L2-norm of the difference between the solution and its counterpart obtained under reflection transformation. The onset of chaos under increasing $\varphi$ is corroborated by the fact that the maximal Lyapunov exponent in Fig.~\ref{fig5}(b) first exhibits a non-negligible positive value $\lambda_{max}=1.4\times 10^{-4}$ at $\varphi=2.13$. Note that in contrast to the onset of a localized extensive chaos, typical for chimeras or bumps, where a certain subset of units unlocks from the coherent background, the transition to chaos here is a collective instability in the sense that all the units within an array immediately exhibit chaotic behavior while the spiking frequencies remain locked within the respective patches. Above the transition, the emerging chaotic patterns do not immediately involve interface units and still feature the piecewise-constant profile of local average spiking frequencies, see Fig.~\ref{fig5}(d). The creation of chaotic patterns featuring interface units with frequencies in-between those of majority and minority patches is rather associated with the reappearance of chaos around $\varphi\approx 2.169$ following a period-four window. In terms of $\omega_k$ profiles, this transition may be understood as a “slope bifurcation” of patched patterns' spatial frequency profile where the sharp transition between the majority and minority patches is replaced by a smoother one, see Fig.~\ref{fig5}(e). Further increasing $\varphi$, the chaotic patterns gain complexity due to a growing number of turbulent interface units, showing the alternating dynamics described in Sec. \ref{switch}, cf. Fig.~\ref{fig5}(f). Meanwhile, the corresponding maximal Lyapunov exponent remains approximately constant, cf. Fig.~\ref{fig5}(b).

\begin{figure}[ht]
\centering
\includegraphics[scale=0.27]{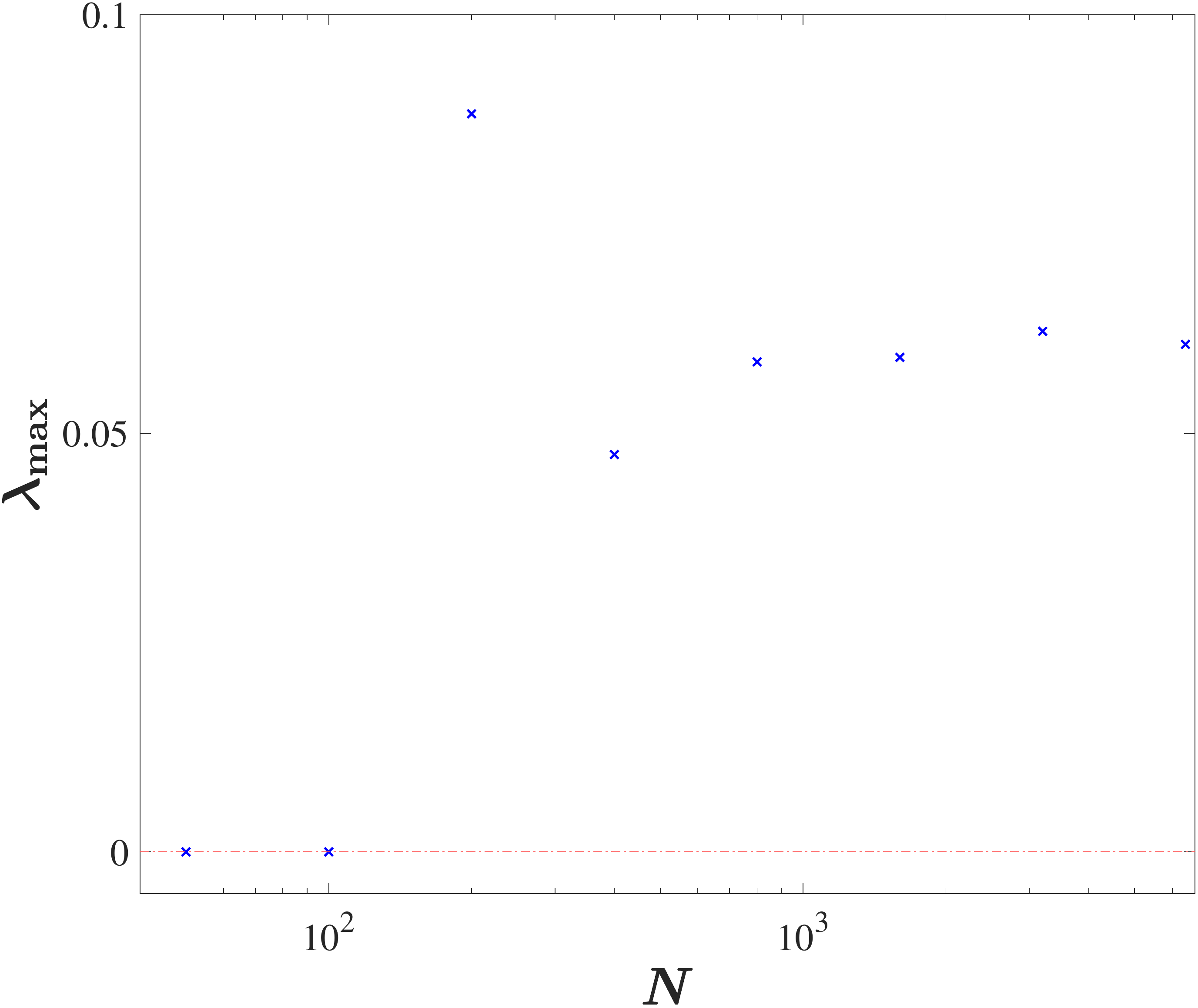}
\caption{Dependence of maximal Lyapunov exponent with system size $\lambda_{max}(N)$. Note the convergence to a finite value $\lambda_{max}\approx 0.06$ for large $N$. Parameters: $\varphi=2.13$, and the remaining ones are the same as in Fig.~\ref{fig1}.}
\label{fig7}
\end{figure}

Next we address the two issues concerning how the system dynamics varies with system size. In particular, we first consider whether and how the observed sequence of bifurcations to chaos depends on $N$, and then examine how the complexity of the solutions changes with $N$. In reply to the former, one notes that for the given coupling strength $\kappa$ and range $R$, the described bifurcation route to chaos qualitatively does not change with $N$ when the initial periodic pattern is constructed by replicating the initial solution for $N=100$. Nevertheless, our simulations indicate that the $\varphi$ values where the particular bifurcations take place shift with increasing $N$, and the threshold for the emergence of chaos apparently reduces with system size. As for the solution complexity, one observes that the number of turbulent interface units grows with $N$ when all the remaining parameters are kept fixed. Finally, chaos is found to persist with increasing $N$, as corroborated by the dependence of the maximal Lyapunov exponent $\lambda_{max}$ on system size in Fig.~\ref{fig7}. There, the coupling parameter is fixed to $\varphi=2.13$, the value just above the transition to chaos for the system size $N=100$, see Figs. \ref{fig5}(a) and \ref{fig6}. As expected, $\lambda_{max}$ for $N=100$ is quite small, but the values calculated for the corresponding solutions at larger $N$ indicate a convergence to a finite value $\lambda_{max}\approx 0.06$ with increasing system size. This is distinct from the classical result for chimeras \cite{WOYM11}, where the maximal Lyapunov exponent decays as $N^{-1/2}$.

\section{Pattern depinning and diffusion of interfaces} \label{diff}

\begin{figure}[ht]
\centering
\includegraphics[scale=0.174]{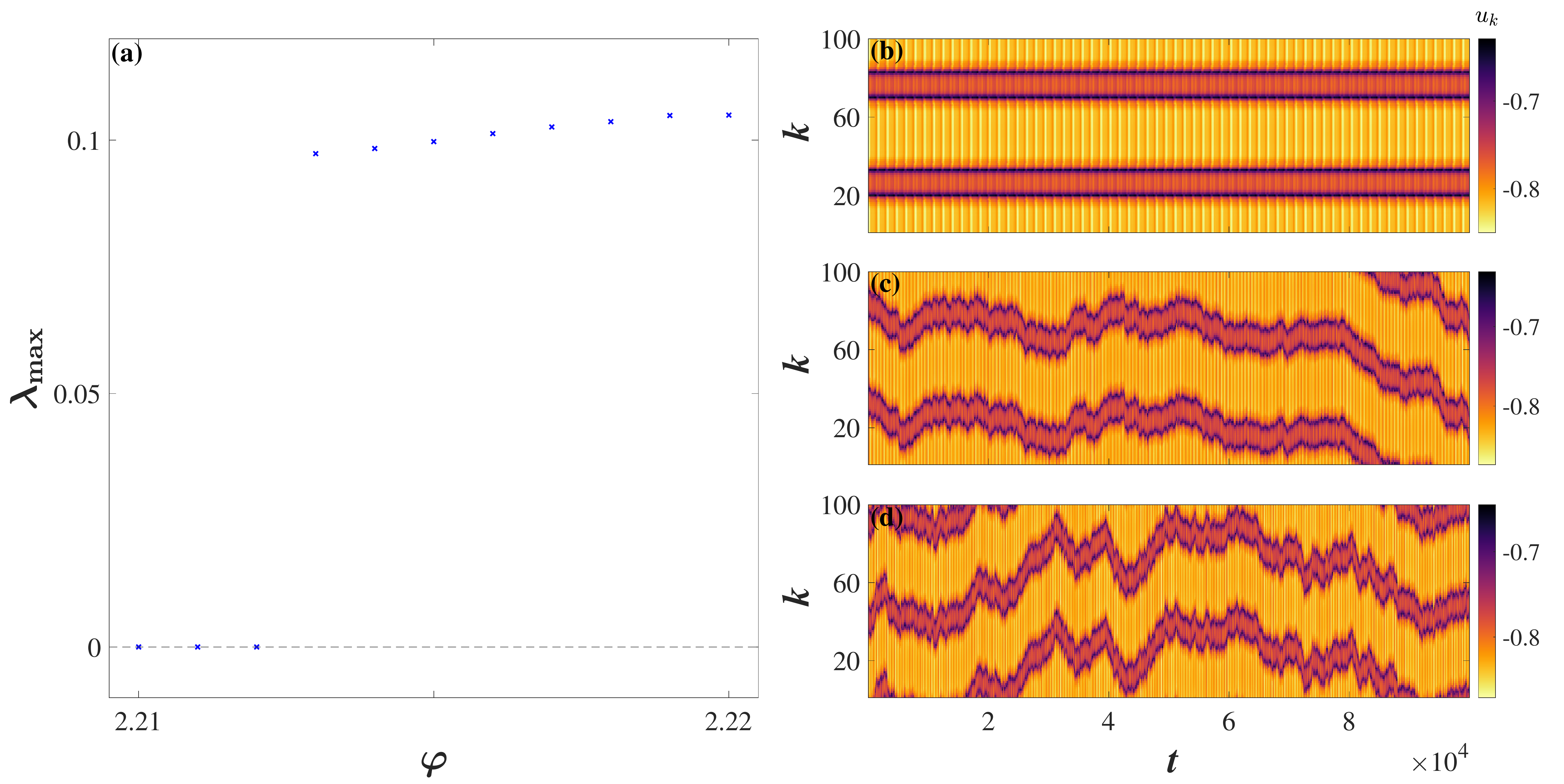}
\caption{Emergence of chaos and diffusion of interfaces for a wavenumber-2 pattern. (a) Dependence of maximal Lyapunov exponent with $\varphi$ indicates onset of chaos for $\varphi\approx 2.213$. (b)-(d) Spatiotemporal evolution of $u_k(t)$ for a periodic pattern at $\varphi=2.21$ and chaotic patterns at $\varphi=2.22$ and $\varphi=2.23$, respectively. System parameters: $R=40,\kappa=0.4,a=1.01,\varepsilon=0.05,N=100$.}
\label{fig8}
\end{figure}

So far, we have considered only static patched patterns which undergo the transition to chaos that is not accompanied by an immediate onset of turbulent interfaces. While this is typical for smaller and intermediate coupling ranges, one finds a rather different scenario if the coupling range $R$ is further increased. Increasing the coupling range affects the primary pattern by reducing its wavenumber, similar to \cite{OOHS13,OZHSS15}. For such patterns, the transition to chaos coincides with the formation of turbulent interfaces, which moreover engage in lateral diffusive motion, similar to the random walk of the incoherent part of chimeras. As an example of this scenario, we have considered the onset of chaos for a periodic patched pattern with the wavenumber two, a solution analogous to that in Fig.~\ref{fig1}(a), but obtained for $R=40$ with all the other parameters preserved. About $\varphi\approx 2.213$, one observes the transition to chaos, as corroborated by the dependence of the maximal Lyapunov exponent with $\varphi$, see Fig.~\ref{fig8}(a). Below the transition, there is just a static periodic pattern, illustrated in Fig.~\ref{fig8}(b) by the spatiotemporal evolution $u_k(t)$, plotting its local time averages within windows of 100 t.u. over a long time horizon of $10^6$ t.u. In contrast to the scenario described in Sec. \ref{chaos}, the interfaces emerge immediately at the transition, and instead of being pinned to the neighboring patches, rather display a Brownian-like motion, see Fig.~\ref{fig8}(c). Due to this, just like in case of chimeras \cite{AB15}, the spatial profiles of average spiking frequencies for such diffusive patched patterns should be flat when considered over sufficiently long time intervals. The diffusive motion becomes more pronounced as $\varphi$ is further increased, cf. Fig.~\ref{fig8}(d).

\begin{figure}[ht]
\centering
\includegraphics[scale=0.172]{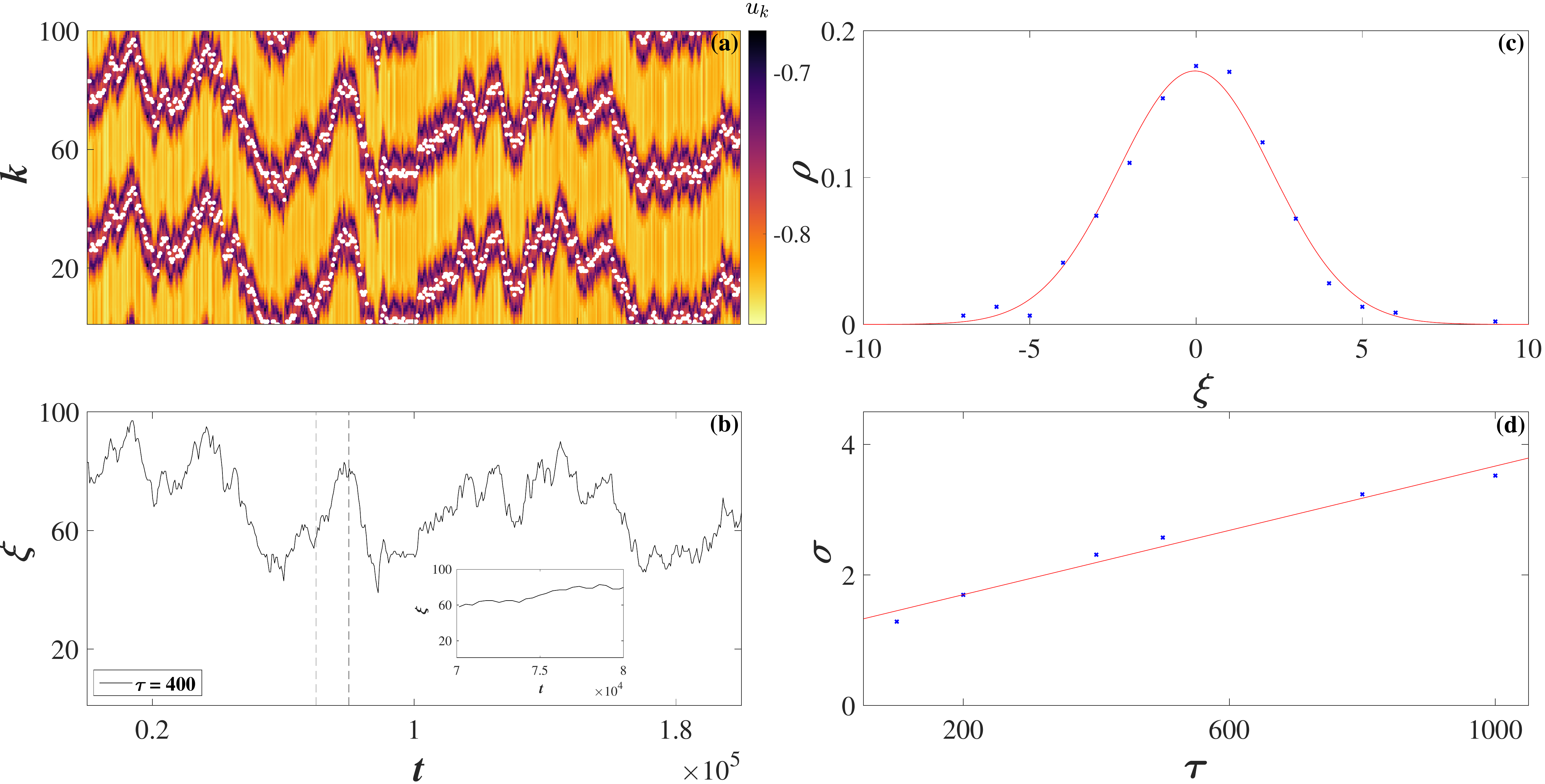}
\caption{(a) white dots: position of the pattern at every time step $\tau=400$ t.u; (b) shifts in position of a single incoherent region bounded by interfaces $\xi(t)$ over a long time horizon of $10^6$ t.u. (main frame) and over a short timescale (inset); (c) Fit of a histogram of displacements $\Delta \xi$ to a Gaussian distribution for $\tau=400$; (d) Variance of Gaussian distributions $\sigma$ as a function of time step $\tau$. System parameters: $\varphi=2.23, R=40,\kappa=0.4,a=1.01,\varepsilon=0.05,N=100$.}
\label{fig9}
\end{figure}

To demonstrate that the motion of interfaces indeed conforms to a Brownian one, we explicitly show that the mean square displacement of the pattern position for the example in Fig.~\ref{fig8}(d) grows linearly with time. The position of the pattern at the given moment is determined following the procedure described in \cite{OWM10}, which essentially entails comparing the vector $u_k(t), k\in[1,N]$ to a suitably chosen periodic reference function $f(x,\xi)$, so that the position of the pattern is given by the $\xi$ value that minimizes the distance between $u_k(t)$ and the reference profile. The results of the procedure are shown in Fig.~\ref{fig9}(a), where white dots, indicating the pattern position $\xi(t)$ after every $\tau=400$ t.u, are plotted on top of the $u_k(t)$ heatmap. Note that the local time averages $\xi(t)$ are used to eliminate fast oscillations. In Fig.~\ref{fig9}(b) are extracted the long-term (main frame) and short-term (inset) motion of a single incoherent region, corresponding to a minority patch of the primary pattern bounded on both sides by the turbulent interface units. Similar to chimeras, the motion of interfaces providing the boundary of a minority region remains correlated, such that the domain does not grow or shrink with time. For a fixed sufficiently large time step $\tau$, the histogram of the corresponding shifts in position $\Delta \xi$ can readily be fitted to a Gaussian distribution
\begin{align}
p(\Delta \xi)&=\frac{1}{\sqrt{2\pi\sigma(\tau)}}e^{-\frac{\Delta \xi^2}{2\sigma(\tau)}},
\end{align}
see Fig.~\ref{fig9}(c) for the case $\tau=400$. Extracting in this way the variances $\sigma (\tau)$ for several values of $\tau$, we demonstrate that they indeed follow a linear dependence of the form $\sigma (\tau)=2D \tau$, see Fig.~\ref{fig9}(d), which can be used to determine the diffusion coefficient $D\approx 1.4\times 10^{-4}$ of the corresponding Fokker-Planck equation.

\section{Summary and discussion}

We have presented patched patterns as a new class of self-organized patterns in coupled excitable systems with nonlocal attractive and repulsive interactions. Pattern formation involves a symmetry breaking, where an assembly of identical units with symmetrical interactions spontaneously splits into patches of frequency locked units, with the majority and minority patches displaying a 1:2 frequency resonance. We have demonstrated that in terms of temporal organization, patched patterns can be classified as periodic, quasiperiodic or chaotic, whereby the former two are always static, while the latter may also show lateral diffusive motion. Apart from patches, chaotic patterns may also include interface units showing chaotic itinerancy, characterized by a slow alternating activity between laminar epochs, where the unit is approximately locked to either of the neighboring patches, and turbulent epochs, with a fast switching between the orbits resembling the local dynamics within the patches. We have explained the basic mechanism of self-organization of patched patterns as an interplay between self-localized excitations and propagation of excitations, the two phenomena classically observed in coupled excitable systems. Using standard chaos and bifurcation theory in finite-dimensional systems, we have disclosed the typical transition route from periodic solutions to chaos via quasiperiodicity as the coupling parameter $\varphi$ is increased. There, chaos emerges from the torus breakup, and the onset of turbulent interfaces is associated with a secondary bifurcation. Nevertheless, the transition to chaos is further found to depend on the wavenumber of the primary pattern, which can be controlled by the coupling range. For sufficiently large coupling ranges admitting wavenumber-2 patterns, we have identified the second scenario of transition to chaos, where its emergence coincides with the formation of diffusive interfaces, explicitly shown to exhibit Brownian-like motion.

Patched patterns we have discovered bear certain resemblance to coherence-incoherence patterns observed so far in coupled oscillators or coupled excitable systems, but also display considerable differences. In particular, patched patterns are different than bumps \cite{L16,LO20,FOW21,L11} because there extensive chaos is spatially localized and the bulk units are stationary (inactive). Also, our patched patterns with interfaces are distinct from classical solitary states because the interface units are not isolated and randomly distributed, but rather form a spatially continuous profile. Distinct from classical chimeras \cite{WOYM11}, maximal Lyapunov exponent for the patched patterns converges to a finite value instead of decaying with the system size. Still, we note a certain similarity to some of the less conspicuous types of patterns observed in coupled oscillators. First, we recall the so-called chimera Ising walls in non-locally coupled Ginzburg-Landau oscillators with a parametric forcing \cite{K07}. There, the incoherent units also form interfaces connecting frequency-locked domains, but in contrast to our patched patterns, the domains at two sides of an interface are 1:1 frequency locked. Second, our class of solutions may be compared to oscillons \cite{SA20}, which also involve a temporally modulated localized spiking activity, as in our minority patches, but such an activity is embedded on an inactive rather than a spiking background. The emergence of spatially incoherent interfaces has also been observed for the so-called mosaic or skeleton patterns in coupled maps \cite{OOHS13,OZHSS15}, but the onset of spatial incoherence there is not associated with temporal chaos in local dynamics. We note that the onset of an alternating activity similar to our interface units has been found for the so-called itinerant chimeras \cite{KKN19}. While this is also not a finite-size effect, it involves all the units within an array, rather than remaining spatially localized. Finally, a recent paper on theta-neuron oscillators mentions non-stationary patterns with the frequency profile similar to ours \cite{OL22}, but instead of spiking, the majority units there are in the state of oscillation death.

The relation between the patched patterns and other types of coherence-incoherence patterns along the path from complete coherence to incoherence in coupled excitable systems requires further study. So far, there is only a partial result suggesting that the patched patterns coexist with solitary states in non-locally coupled arrays of FitzHugh-Nagumo units \cite{FESZ22}, and that the noise promotes patched patterns at the expense of solitary states. The presented results, together with \cite{FOW21,FESZ22}, indicate that the study of self-organized coherence-incoherence patterns in coupled excitable systems opens up interesting new directions of research, revealing types of solutions that bear only a partial resemblance to those in coupled oscillators. An interesting problem would be to investigate these new types of solutions for models amenable to a rigorous analysis of the system behavior in the continuum limit.

\begin{acknowledgments}
I.F. acknowledges funding from Institute of Physics Belgrade through grant by Ministry of Education, Science and Technological Development of Republic of Serbia.
\end{acknowledgments}

\section*{Data Availability}
The data that support the findings of this study are available from the corresponding author upon reasonable request.

\end{document}